\documentclass[codesnippet]{jss}
\usepackage{amsmath} 
\usepackage{verbatim} 

\author{Alireza S. Mahani\\Scienctific Computing Group\\Sentrana Inc. \And 
        Mansour T.A. Sharabiani\\National Heart and Lung Institute\\Imperial College London}
\title{Efficient SIMD RNG for Varying-Parameter Streams: \proglang{C++} Class \code{BatchRNG}}

\Plainauthor{Alireza Mahani, Mansour Sharabiani} 
\Plaintitle{Efficient SIMD RNG for Varying-Parameter Streams: C++ Class BatchRNG} 
\Shorttitle{C++ Class BatchRNG} 

\Abstract{
Single-Instruction, Multiple-Data (SIMD) random number generators (RNGs) take advantage of vector units to offer significant performance gain over non-vectorized libraries, but they often rely on batch production of deviates from distributions with fixed parameters. In many statistical applications such as Gibbs sampling, parameters of sampled distributions change from one iteration to the next, requiring that random deviates be generated one-at-a-time. This situation can render vectorized RNGs inefficient, and even inferior to their scalar counterparts. The \proglang{C++} class \code{BatchRNG} uses buffers of base distributions such uniform, Gaussian and exponential to take advantage of vector units while allowing for sequences of deviates to be generated with varying parameters. These small buffers are consumed and replenished as needed during a program execution. Performance tests using Intel Vector Statistical Library (VSL) on various probability distributions illustrates the effectiveness of the proposed batching strategy.
}
\Plainkeywords{gibb sampling, bayesian, monte carlo markov chain} 

\Address{
  Alireza S. Mahani\\
  Scientific Computing Group\\
  Sentrana Inc.\\
  Washington DC, 20006\\
  E-mail: \email{alireza.mahani@sentrana.com}\\
}

\begin{document}


\section{Introduction}\label{sec-intro}
Vectorized random number generators (RNGs) such as Intel MKL Vector Statistical Library\footnote{\url{https://software.intel.com/en-us/mkl_11.1_vslnotes}}, SIMD-oriented Fast Mersenne Twister (SFMT) (\cite{saito2008simd}) and Nvidia's cuRAND\footnote{\url{https://developer.nvidia.com/cuRAND}} offer significant speedup over non-vectorized RNGs due to their utilization of Single-Instruction, Multiple-Data (SIMD) machine instructions\footnote{See Chapter 4 of \cite{hennessy2012computer} for an excellent overview of SIMD architectures.} for parallel execution of bitwise, integer and floating-point operations. To vectorize efficiently, however, these libraries typically require that a large number of random numbers (typically more than 1000) are generated from a given distribution with fixed parameters (batch scenario). For example, below is the VSL function for generating \code{n} double-precision deviates from a Uniform distribution with left- and right-boundary parameters \code{a} and \code{b}, respectively:

\code{vdRngUniform(method, stream, n, r, a, b);}

The output is written to a double buffer of length \code{N}, starting at memory address \code{r}. On our test machine\footnote{See Section~\ref{subsec-setup} for details of experimental setup.}, we measured the performance of \code{vdRngUniform} to be around 4.3 Clock Cycles Per Sample (CCPS).

In many statistical applications, unfortunately, such batch scenarios may not exist. In particular, as we transition from pure Monte Carlo to Monte Carlo Markov Chain (MCMC) simulations, the Markovian nature of the process requires streams of random numbers to be drawn from distributions whose parameters change from one iteration to the next. We illustrate the point with a simple example.

Imagine a 2-dimensional probability density function (PDF) defined as a uniform density over the triangle with coordinates (0,0), (0,1) and (1,0). The PDF is formally defined as:
\begin{equation} \label{eq-2d-uniform}
P(X,Y) = \begin{cases} 1/2 & \mbox{If } X>0 \mbox{ \& } Y>0 \mbox{ \& } X+Y<1, \\ 0 & \mbox{otherwise.} \end{cases}
\end{equation}
This joint PDF is not separable, i.e. $P(X,Y) \neq P(X) \times P(Y)$. To draw samples from it we can use Gibbs sampling (\cite{christopher2006pattern}), where we alternate between $X$ and $Y$, drawing a sample from each conditional distribution, given the last sampled value for the other coordinate\footnote{The integrals involving this joint distribution are simple enough to be analytically tractable, thus not requiring the application of Gibbs sampling. The example serves a pedagogical purpose, nonetheless.}. The code snippet in Figure~\ref{fig:2d-uniform-gibbs} implements Gibbs sampling for the above distribution. We see that, due to Markovian dependence among successive iterations of the \code{for} loop, random numbers must be generated one at a time (OAAT scenario). This has a drastic negative impact on performance, producing a CCPS of 65.6, i.e. a 15.3x drop from the Batch scenario.

\begin{figure}
\begin{verbatim}
--------------------------------------------------------------------------
double x=0.5, y=0.5;
for (int i=0; i<n; i++) {
  vdRngUniform(method, stream, 1, &x, 0.0, 1.0-y);
  vdRngUniform(method, stream, 1, &y, 0.0, 1.0-x);
}
--------------------------------------------------------------------------
\end{verbatim}
\caption[]{Gibbs sampling of the joint PDF described in Equation \ref{eq-2d-uniform}. Initialization of \code{method} and \code{stream} variables, as well as saving of samples to arrays for further analysis, are not shown for brevity.}
\label{fig:2d-uniform-gibbs}
\end{figure}


While the simple example presented above is contrived, it illustrates a concept that appears in many real-world applications. For example, in Gibbs sampling of Latent Dirichlet Allocation (LDA) models for topic analysis (\cite{blei2012probabilistic}), one must sample from a Dirichlet distribution whose parameters are updated during each iteration. Dirichlet deviates can easily be generated from Gamma deviates\footnote{For a K-dimensional Dirichlet deviate with parameter $\mathbf{\alpha}$, we first draw K $y_i$'s from $\mathrm{Gamma}(\alpha_i,1)$. Next we normalize: $x_i = y_i/\sum_i y_i$. $\mathbf{x}$ is the desired Dirichlet deviate.}. Analyzing the performance of Intel VSL's Gamma RNG function \code{vdRngGamma} shows a similar performance degradation from Batch to OAAT scenario (see Figure~\ref{fig-results}).

This performance drop of vectorized RNG libraries from Batch to OAAT scenarios is by no means surprising since overhead of data movement and arithmetic instructions dominates performance for short vectors. The case of \code{n=1} is an extremely suboptimal scenario for vectorized RNG functions. The \proglang{C++} class \code{BatchRNG} is designed to address this performance loss of vectorized RNG libraries for generating varying-parameter RNG streams.


\section{C++ Class BatchRNG}\label{sec-batchrng-class}
In this section, we review the key concepts behind the \proglang{C++} \code{BatchRNG} class (Section~\ref{subsec-key-concepts}), present salient implementation aspects of the class (Section~\ref{subsec-implement}), review experimental setup (Section~\ref{subsec-setup}), and discuss performance results (Section~\ref{subsec-performance}).

\subsection{Key Concepts}\label{subsec-key-concepts}
\code{BatchRNG} relies on two concepts regarding random number generation:
\begin{enumerate}
\item Computer algorithms for generating random deviates for all distributions rely on the same base algorithm: generating a random number (integer or floating-point) from a uniform distribution. Some distributions such as Gamma rely on uniform and Gaussian RNGs, but the latter itself relies on uniform deviates.
\item For a subset of probability distributions including uniform, Gaussian, and exponential, samples from a standard distribution with fixed parameters can be easily (often linearly) transformed to generate deviates for the same family, and with arbitrary parameters. For example, a random deviate \code{x} from the standard uniform distribution \code{U[0,1)} can be turned into a random deviate from \code{U[a,b)} by the linear transformation: \code{a + (b - a) * x}. We call such distributions `reducible distributions'.
\end{enumerate}

Based on the above two points, \code{BatchRNG} creates and maintains buffers of random deviates for reducible distributions (uniform, Gaussian, exponential) as well as their transformations used in other distributions (e.g. logarithm of uniform deviates). These buffers can be used to 1) generate random deviates of arbitrary parameters for those same reducible distributions, and 2) feed the algorithms used to generate deviates for irreducible distributions.

\subsection{Implementation}\label{subsec-implement}
The class \code{BatchRNG} assumes the existence of at least a vectorized RNG for uniform distribution. Vectorized functions for other reducible distributions such as Gaussian and exponential are not functionally necessary but they significantly improve performance results if they are efficiently vectorized. In addition to these core routine(s), most RNG libraries include allocation and de-allocation routines, which are wrapped in \code{BatchRNG}'s constructor and destructor.

\code{BatchRNG} contains the following class members:
\begin{itemize}
\item Data structure for the core engine (e.g. \code{VSLStreamStatePtr}).
\item \code{buffer\_length} integer fields, representing the length of double-precision buffers for reducible distributions.
\item Double-precision \code{buffer} pointers to heap memory addresses containing reducible-distribution batches.
\item \code{buffer\_counter} integer fields, holding index of next random deviate for each reducible distribution that must be fetched.
\item Any \code{typedef}'s used by the core engine to identify the methods used for each reducible distribution.
\end{itemize}

\code{BatchRNG} contains the following methods:
\begin{itemize}
\item Class constructor: In addition to calling the core engine allocator, it allocates memory for all reducible-distribution buffers and sets their corresponding counters to zero, i.e. the beginning of the buffers. Finally, the constructor fills the buffers by calling the corresponding RNG function of the core engine.
\item Class destructor: In addition to calling the core engine de-allocator, it frees heap memory allocated by the constructor for reducible-distribution buffers.
\item RNG functions:
\begin{itemize}
\item Rreducible Distributions: First, it checks the buffer counter to see if it is pointing to the end of buffer. If yes, it calls core engine RNGs to re-fill the buffer, and it resets the counter. Current implementation includes functions \code{GetUniform}, \code{GetGaussian} and \code{GetExponential} as well as \code{log} of uniform distributions: \code{GetLogUniform}.
\item Irreducible Distributions: The implementation utilizes the reducible-distribution functions discussed above. Current implementation contains \code{GetLaplace}, \code{GetWeibull} and \code{GetGamma}.
\end{itemize}
\end{itemize}

\subsection{Experimental Setup}\label{subsec-setup}
\textbf{Hardware} All tests are conducted on an Intel Xeon E5-2670\footnote{\url{http://ark.intel.com/products/64595/}} (2.6GHz, 8 cores per socket). Cache sizes are 32KB/256KB/20MB for L1/L2/L3.

\textbf{Software} Current implementation of \code{BatchRNG} uses the Intel MKL Vector Statistical Library as core engine. Intel C++ compiler is used to compile all source files (class and test files). All Intel software is part of Intel Composer XE 2013 suite. Optimization flag \code{-O3} was used throughout the tests.

\textbf{List of Methods Compared} Here we describe each of the columns for which performance results are reported in Figure~\ref{fig-results}:
\begin{itemize}
\item \textbf{VSL-Batch}: Calling the Intel VSL function in Batch mode, i.e. a single call with fixed parameters.
\item \textbf{VSL-OAAT}: Calling the Intel VSL function inside a \code{for} loop, i.e. One-at-A-time mode. Distribution parameters are read as elements of an array in memory. See "Methodological Details".
\item \textbf{BatchRNG}: Calling the \code{BatchRNG} functions inside a \code{for} loop in a OAAT scenario. Distribution parameters are read as elements of an array in memory. See "Methodological Details".
\end{itemize}
The "speedup" column in Figure~\ref{fig-results} shows the performance gain of \code{BatchRNG} over VSL-OAAT.

\textbf{Methodological Details} Measuring performance of random number generation routines involves several subtleties. The actual time spent on generating random numbers consists of, not only the computation inside RNG functions, but also the data movement between memory, cache and registers. The latter component, in turn, depends on the program context in which RNG functions are called: Are the parameters of the distribution cached or must they be fetched from memory? Are the RNG's being written to a buffer or consumed as they are generated? If we are writing to a buffer, is the  compiler generating instructions for reading the buffer from memory? To avoid getting lost in such details, we note two objectives of our performance tests: 1) We are less interested in absolute performance of different methods, and more interested in their relative performance, 2) We care less about exact numbers and small differences, and more about general patterns and large differences. With these two principals in mind, below we outline some of our methodological choices and a brief discussion of their relevance and impact on performance measurements.
\begin{enumerate}
\item In all RNG methods listed above, we write output random numbers to a memory buffer. This incurs one memory write operation per random number generated, thus increasing the CCPS numbers. In real-world applications, such writes may not be needed but we chose to include this to facilitate calculation of RNG statistics (mean/std) and also to prevent the compiler from optimizing away the random number generation code.
\item The memory buffer holding the output RNG stream is cleared before RNG function is called. This allows the compiler to optimize away an unnecessary memory read operation per RNG cycle.
\item For OAAT scenarios, we read the parameters from a memory buffer, which imposes a memory read operation per random number generated. In real-world applications, distribution parameters may be cached when RNG function is called.
\end{enumerate}

\subsection{Performance Results}\label{subsec-performance}
Figure \ref{fig-results} shows the performance of \code{BatchRNG} (OAAT scenario) in comparison to Intel VSL, applied in Batch and OAAT scenarios. (See Section~\ref{subsec-setup} for detailed definition of each column.)

\begin{figure}
\caption{Performance comparison of \code{BatchRNG} against . Performance is measured in Clock Cycles Per Sampler or CCPS. Numbers are averages for 200M random samples generated for each distribution and method.}
\label{fig-results}
\resizebox{\textwidth}{!}{
\begin{tabular}{|l|l|l|l|l|}
\hline
Distribution & VSL-Batch & VSL-OAAT & Batch-RNG & Speedup \\
\hline
Uniform & 4.3 & 65.1 & 12.4 & 5.2x \\
\hline
Gaussian & 25.5 & 1105.4 & 34.6 & 31.9x \\
\hline
Exponential & 12.2 & 646.0 & 14.2 & 45.5x \\
\hline
Laplace & 15.7 & 706.4 & 41.9 & 16.9x \\
\hline
Weibull & 16.3 & 857.6 & 22.9 & 37.5x \\
\hline
Gamma & 51.0 & 1316.0 & 84.0 & 15.7x \\
\hline
\end{tabular}
}
\end{figure}

We see that \code{BatchRNG} is 4x-45x faster than Intel VSL running in OAAT mode, in some cases coming very close to the VSL-Batch result (e.g. Exponential and Weibull). Two major factors are responsible for the performance gap between VSL-Batch and \code{BatchRNG}:
\begin{itemize}
\item Buffer management: a) checking buffer counter, b) fetching random number from buffer, c) incrementing buffer counter, d) resetting buffer counter after replenishment. This is the sole factor responsible for Uniform, Gaussian, and Exponential results.
\item Incomplete vectorization: \code{BatchRNG} only vectorizes generation of reducible-distributions deviates by calling the core engine with vectorized functions. However, a fully-vectorized RNG function must also include vectorizaion of arithmetic/transcendental operations involved. This is why having more vectorized functions in the core RNG engine boosts performance. Even if we could vectorize the math operations inside a function, the efficiency would not be the same as a natively-vectorized function since the latter could pipeline several vectorized operations, leading to more efficiency.
\end{itemize}

\code{BatchRNG} constructor expects a set of parameters that specify the length of reducible-distribution buffers. We have set a default value of 1000 for all these buffers. Choosing too small a buffer leads to inefficient vectorization inside the core RNG functions, while a large buffer can lead to wasted CPU cycles on random numbers that are not used by the end of program run. We find 1000 to be a good number in that it offers most of the vectorization gain while imposing negligible waste overhead in worst-case scenarios. For example, generating 1000 samples from a Gamma distribution would only take 84,000 clock cycles on our test server, which translates into 32usec. This is the upper-bound on wasted time due to incomplete buffer usage, which is acceptable in most applications. Another possible optimization is to use a buffer length of zero for those reducible distributions that are not useful in a particular application. For example, if one only needs uniform and Gaussian RNGs in an application, then the buffer lengths for exponential and log-uniform can be set to zero. Again, this would only save a negligible amount of time.

\section{Summary and Future Work}\label{sec-discuss}
We presented \code{BatchRNG}, a \proglang{C++} class to improve the performance of vectorized RNG libraries for producing varying-parameter random deviates, with application in MCMC techniques such as Gibbs sampling of high-dimensional posterior distributions in Bayesian statistics and machine learning models. Using Intel VSL as core engine, we showed that \code{BatchRNG} leads to significant performance improvement over a range of statistical distributions.

There are several improvements and new features to be considered for future releases of \code{BatchRNG}:
\begin{enumerate}
\item Expanded library: Current implementation has a core set of distributions, but this set must definitely be expanded to include more continuous distributions as well as important discrete distributions.
\item Alternative Core RNGs: Current implementation uses Intel VSL - a proprietary software - as its core engine. Future work can add alternative vectorized RNG software such as SIMD-oriented Fast Mersenne Twister (SFMT) (\cite{saito2008simd}). (The current version of this open-source software does not support AVX vector instructions.)
\item \code{BatchRNG} as asynchronous service: In current implementation, RNG functions such as \code{GetUniform} check buffer counter and kick off replenishment process when needed. This can be considered a blocking call to the batch RNG service. It is conceptually possible to improve performance by creating a batch service that runs in the background, using spare CPU cycles or in overlap with other computational work, to replenish the RNG buffers. The added complexity of the software must be justified by real-world use cases where such optimizations would make a significant impact on overall application performance.
\item Parallel Co-processors: The same batching strategy can be adapted for parallel co-processors such as Graphic Processing Units (GPUs) and Intel Xeon Phi. Justification and development strategy for such software must be done in the context of a particular class of applications. An important factor is whether/how much of the application is offloaded to the coprocessor. For example, in the extreme case where we want to use the coprocessor only as an RNG engine, the memory bandwidth limit (using PCI Express) between the processor and coprocessor may impose such a heavy overhead that any speedup from offloading RNG service to the coprocessor is neutralized. Another deciding factor is the opportunity cost of utilizing parallel threads for SIMD random number generation. If there exist better parallelization opportunities in the application, they may be favored over batch RNG on the coprocessor.
\end{enumerate}

Note: Source code for \code{BatchRNG} can be obtained from \url{https://sites.google.com/site/alirezasmahaniphd/home/software/BatchRNG.tar.gz}. (Please note that, due to third-party copyright reasons, we cannot publicly share the \code{GetGamma} function. We are working on developing an open-source version, and will share as soon as available.)

\bibliography{batchrng}

\begin{thebibliography}{4}
\newcommand{\enquote}[1]{``#1''}
\providecommand{\natexlab}[1]{#1}
\providecommand{\url}[1]{\texttt{#1}}
\providecommand{\urlprefix}{URL }
\expandafter\ifx\csname urlstyle\endcsname\relax
  \providecommand{\doi}[1]{doi:\discretionary{}{}{}#1}\else
  \providecommand{\doi}{doi:\discretionary{}{}{}\begingroup
  \urlstyle{rm}\Url}\fi
\providecommand{\eprint}[2][]{\url{#2}}

\bibitem[{Bishop(2006)}]{christopher2006pattern}
Bishop CM (2006).
\newblock \enquote{Pattern Recognition and Machine Learning.}

\bibitem[{Blei(2012)}]{blei2012probabilistic}
Blei DM (2012).
\newblock \enquote{Probabilistic topic models.}
\newblock \emph{Communications of the ACM}, \textbf{55}(4), 77--84.

\bibitem[{Hennessy and Patterson(2012)}]{hennessy2012computer}
Hennessy JL, Patterson DA (2012).
\newblock \emph{Computer architecture: a quantitative approach}.
\newblock Elsevier.

\bibitem[{Saito and Matsumoto(2008)}]{saito2008simd}
Saito M, Matsumoto M (2008).
\newblock \enquote{SIMD-oriented fast Mersenne Twister: a 128-bit pseudorandom
  number generator.}
\newblock In \emph{Monte Carlo and Quasi-Monte Carlo Methods 2006}, pp.
  607--622. Springer.

\end{thebibliography}

\end{document}